\documentclass{article}
\usepackage{arxiv}
\usepackage{gensymb}
\usepackage{graphicx,amsmath,amssymb,amsfonts}
\usepackage{enumerate,setspace}
\usepackage{psfrag}
\usepackage{color}   
\usepackage{rotating}
\usepackage{pbox}
\usepackage{caption}
\usepackage{subcaption}
\usepackage{framed}
\usepackage{siunitx}
\newcommand{\comment}[1]{}
\renewcommand{\comment}[1]{{\noindent \color{blue} #1}}
\newcommand{\commentR}[1]{}

\newcommand{\NN}{{\bf N}}
\newcommand{\YY}{{\bf Y}}
\renewcommand{\AA}{{\bf A}}
\newcommand{\EQ}{\begin{equation}} 
\newcommand{\EE}{\end{equation}}
\newcommand{\EQA}{\begin{eqnarray}}
\newcommand{\EEA}{\end{eqnarray}}

\renewcommand{\|}{\, | \,}

\usepackage{soul}
\usepackage{adjustbox}
\usepackage{authblk}

\title{Predicting \textit{in vivo} escape dynamics of HIV-1 from a broadly neutralizing antibody}

\author[a]{Matthijs Meijers}
\author[b]{Kanika Vanshylla}
\author[b]{Henning Gruell}
\author[b,c,d]{Florian Klein}
\author[a]{Michael L{\"a}ssig}

\affil[a]{Institut f\"ur Biologische Physik, Universit\"at zu K\"oln, 50937, Cologne, Germany}
\affil[b]{Laboratory of Experimental Immunology, Institute of Virology, Faculty of Medicine and University Hospital Cologne, University of Cologne, 50931 Cologne, Germany}
\affil[c]{German Center for Infection Research (DZIF), Partner Site Bonn Cologne, 50931 Cologne, Germany}
\affil[d]{Center for Molecular Medicine (CMMC), University of Cologne, Cologne, Germany}

\begin{document}

\maketitle

\begin{abstract}
Broadly neutralizing antibodies are promising candidates for treatment and prevention of HIV-1 infections.
Such antibodies can temporarily suppress viral load in infected individuals; however, the virus often rebounds by escape mutants that have evolved resistance. In this paper, we map an {\em in vivo} fitness landscape of HIV-1 interacting with broadly neutralizing antibodies, using data from a recent clinical trial. We identify two  fitness factors, antibody dosage and viral load, that determine viral reproduction rates reproducibly across different hosts. The model successfully predicts the escape dynamics of HIV-1 in the course of an antibody treatment, including a characteristic frequency turnover between sensitive and resistant strains. This turnover is governed by a dosage-dependent fitness ranking, resulting from an evolutionary tradeoff between antibody resistance and its collateral cost in drug-free growth. Our analysis suggests resistance-cost tradeoff curves as a measure of antibody performance in the presence of resistance evolution. 
\end{abstract}

HIV-1 infection is characterized by a high turn-over rate in combination with high mutation rates, which contribute to the virus' extraordinary capacity to evade the immune response of the host \cite{Wei1995}. 
In return, the immune system explores a broad set of responses, creating a co-evolution of the two systems \cite{Liao2013,Nourmohammad2016}. 
In recent years, the discovery of broadly neutralizing antibodies (bnAbs) created the prospect of antibody-mediated prevention \cite{Pegu2017, Burton2016} or treatment  \cite{Lynch2015a, Bar2016, Mendoza2018} of HIV-1 infection. 
Such antibodies have been isolated from a minority of HIV-1-infected individuals, termed elite neutralisers, and can potently neutralize a large spectrum of existing HIV-1 strains. They target relatively conserved sites on the envelope protein (Env) of the virus~\cite{Klein2013}, blocking virus entry into cells. Additionally, they can engage the host immune system to target infected cells \cite{Lu2016}. 
However, clinical studies show that in individuals infected with sensitive viral strains, antibody treatment leads to an initial decline in viral load, which is followed by a rebound on a time scale of weeks~\cite{Caskey2015, Caskey2017, Lynch2015}. Specific insight into the underlying viral escape from antibody neutralization comes from a recent study recording viral populations in a cohort of HIV-1 infected individuals following infusion of a specific bnAb~\cite{Caskey2017}.  This study provides time-resolved, {\em in vivo} data including  viral load, antibody concentration, and single-virion genome sequences. It reveals a complex escape dynamics involving several resistant mutant strains.   

In this paper, we establish a biophysically grounded, predictive fitness landscape for {\em in vivo} HIV-1 escape dynamics from bnAbs. In the first part of the paper, we use the data of ref.~\cite{Caskey2017} to infer viral within-host replication rates of sensitive and resistant strains. These rates depend on two key factors: the antibody concentration, which determines the likelihood of antibody-antigen binding, and the virion density, which is subject to saturation effects setting the carrying capacity of the HIV-1 population in a given patient. Both of these densities vary by orders of magnitude in the course of a treatment protocol. Remarkably, however, we can quantify their effects on viral growth with a universal fitness model, reflecting dynamical dependences reproducible across individuals. 

Our fitness model contains two key parameters characterizing a given HIV-1 strain: its growth rate in the absence of antibodies and its resistance to the bnAb, defined as the reduction in the antibody-antigen binding that confers neutralization of viral growth. We infer an {\em in vivo} tradeoff between these parameters: mutants with increased antibody resistance have a reduced reproductive rate in the absence of antibodies. This tradeoff reflects a well-known general phenomenon: resistance mutations can involve collateral costs \cite{Sanjuan2010}. Mutations can reduce the stability of protein folding and thereby decrease fitness \cite{Wylie2011, Gong2013}. Moreover, since bnAbs target conserved sites of the virus, mutations in these regions can impair viral function. The cost of escape mutations can be quantified in several ways. Using virus replication assays, Lynch \emph{et al.} showed that escape from VRC01-class bnAbs targeting the CD4 binding site results in reduced viral replication \cite{Lynch2015}. Computational methods based deep mutational scanning \cite{Haddox2016}, deep sequencing of longitudinal samples \cite{Neher2017}, or multiple sequence alignment \cite{Louie2018} can also reveal a fitness cost at target sites. Similar tradeoffs between evolution of resistance and function have been observed in microbial systems \cite{Andersson2010}, and for cancer \cite{Balachandran2017}. However, to our knowledge, a direct inference of the resistance-growth rate tradeoff for \emph{in vivo} escape from bnAbs has not been performed so far. This tradeoff has an important dynamical consequence: it determines a bnAb-dosage-dependent fitness ranking of {sensitive and resistant strains, which, in turn, drives strain turnover during the escape process. 

In the second part of the paper, we show that the fitness model can successfully predict viral escape dynamics during bnAb treatment, given host-specific initial data of viral load and standing variation of resistance mutations prior to bnAb exposure. Predictable features of these dynamics include the strain turnover during the escape process, manifesting the universal, dosage-dependent ranking of strains in our fitness model. We discuss the consequences of our findings for the evolutionary optimization of antibodies and of time-dependent treatment protocols.

\section*{Results}
\paragraph{Time-resolved \emph{in vivo} data of bnAb escape.}
In the study of Caskey \emph{et al.}, 19 individuals infected with HIV-1 received a single infusion of the of the V3-loop directed antibody 10-1074. The total viral load $N$ (measured in RNA copies/ml), and the bnAb dosage $A$ (measured in $\mu$g/ml) were tracked over several weeks after the infusion. Single-virion genome sequencing was performed on plasma samples obtained at specific time points, providing frequency estimates of sensitive and mutant strains. 
Additionally, the neutralizing power of the antibody against the sensitive strain was measured in terms of the IC50 bnAb concentration, $K_{\rm wt}$. Here we infer analogous IC50 values for mutant strains and use them as a measure of bnAb resistance.
We use data from 11 of the individuals studied in ref.~\cite{Caskey2017}; these individuals responded to the bnAb infusion, were not on anti-retroviral therapy, and single-genome sequencing was performed at three or more time points; see SI Appendix for details. Sequence analysis and phenotype analyses with pseudoviruses shows that the viral escape from neutralization can be associated with amino acid changes at one of the gp120 epitope residues 334, 332, or 325 away from the wild-type (wt) allele that is common to all sensitive strains; see also refs.~\cite{Bjorkman2012,Bloom2019}. The escape mutations at residue $334$ or $332$ eliminate a glycosylation site, at which the antibody makes a critical contact to a glycan. Given this functional equivalence, we group them together as mutant 1 (mt1). The escape mutation at residue $325$  (mt2) alters a different contact site of the antibody. We will show that the fitness of the sensitive (wt) and resistant (mt1, mt2) strains is largely independent of the genetic background, which differs between viral populations in different host.

\begin{figure*}[t]
\centering
  \begin{adjustbox}{center}
   \includegraphics[height=100mm,width= 0.95 \textwidth ]{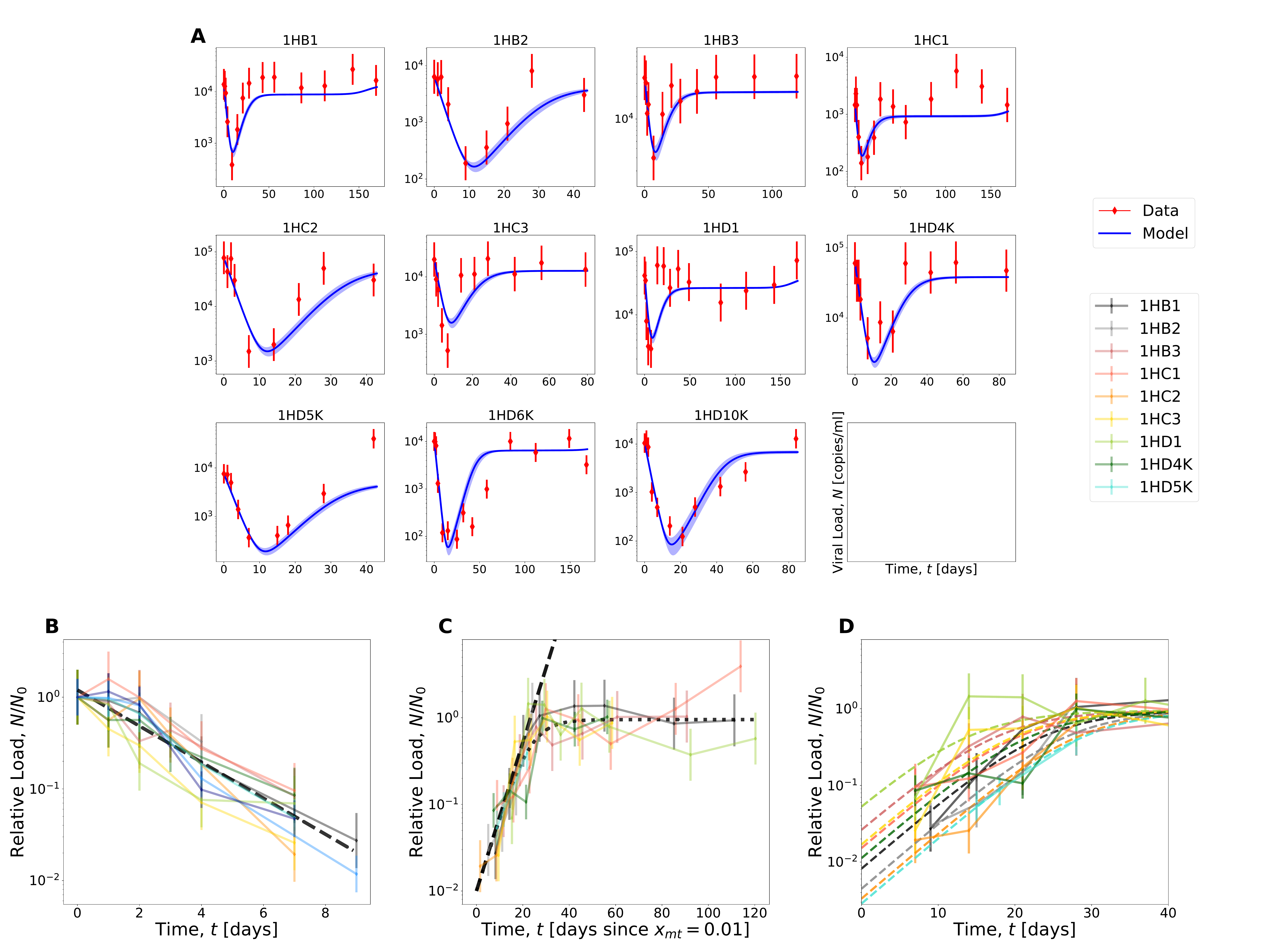}
 \end{adjustbox}
\caption{\small  
{\bf Viral load trajectories have universal growth parameters.} 
(A)~Observed time series of the viral load in 11 individuals  [RNA copies / ml] (red dots) are shown together with the load trajectory of the maximum-likelihood fitness model. 
(B)~Collapse plot of the initial load decline. Measured relative load $N/N_0$ (solid lines) and universal (host-independent) exponential fit with inferred clearance rate $d =  (0.45 \pm 0.03)$ ${\rm day}^{-1}$ (dashed line).
(C)~Collapse plot of the load rebound. Measured relative load $N/N_0$ plotted against the time from a common initial value $N/N_0 = 10^{-2}$ (solid lines), universal fit curve (dotted line), and exponential fit to the initial rebound with inferred mutant growth rate $f^0_{\rm mt1} =  (0.20 \pm 0.04)$ ${\rm day}^{-1}$ (dashed line).
(D)~Inference of the initial mutant frequency. The measured relative load $N/N_0$ is plotted against the time from the start of treatment at $t = t_0$ (solid lines).  Extrapolation of the exponential rebound back to $t_0$ (dashed lines), provides estimates of the initial frequencies $x_{0, {\rm mt1}}$ (intercept with the vertical axis). 
}
\label{fig1}
\end{figure*}

\paragraph{A biophysical fitness model for viral escape.} 
We describe the reproductive dynamics of HIV-1 by a continuous birth-death process. A given viral strain $i$ (wt, mt1, or mt2) has an intra-host replication (birth) rate that depends on the antibody dosage and on an effective viral load, 
\EQ
b_i (A,N_u) = b_i^0 \, \frac{\exp(-CN_u)}{1 + A/K_i}. 
\label{eq:birth_rate}
\EE
In this model, the basic replication rate, $b_i^0$, is reduced by two factors characterizing the intra-host environment. First, functional antibody binding prevents new cell infections, constraining replication to the fraction of unbound virions, $p_u = 1/(1 + A/K_i)$. This fraction depends on the antibody dosage $A$ and the strain-specific 
antibody resistance $K_i$, defined as the dissociation constant of binding that confers neutralization. Similar biophysical fitness landscapes linking functional binding interactions and growth have been established in other microbial and viral systems \cite{Gerland2002, Berg2004, Mustonen2008, Magnus2016, Rotem2018}. 
Second, either immune pressure from cytotoxic cells on infected cells or local depletion of uninfected CD4 T-cells \cite{Boer1998} can decrease the replication rate of the virus by a factor $q_u = \exp(-CN_u)$ (SI Appendix). Here, $C$ is a constraint parameter that sets the host-specific carrying capacity and $N_u = \sum_i N_i / (1 + A/K_i)$ is an effective viral load, defined as the total number of infective virions (which are not neutralized not by antibodies). Neutralized, bnAb-bound virions do not count for these saturation effects, because they do not contribute to cell infection.
Furthermore, we describe clearance of virions, either by active immune processes or by decay, by a single clearance (death) rate $d$, assuming that differences between strains are negligible. Replication and clearance determine the net growth rate or absolute fitness of a given strain, 
\EQ
f_i (A,N_u) = b_i (A,N_u) - d. 
\label{eq:fitness}
\EE
The fitness in the regime of low load, i.e, in the absence of saturation effects, is denoted by the shorthand $f_i(A) = f_i(A, N_u = 0)$.

\paragraph{Model-based eco-evolutionary dynamics.} 
The fitness model~[\ref{eq:fitness}] sets the mutation-selection dynamics of individual viral strains, 
\EQ
\dot{N}_i = f_i (A, N_u) \, N_i  + \mu M_i. 
\label{eq:sub_pop}
\EE
Here the terms $M_i$ describe the mutational turnover between strains given by $M_{\rm mt1} = 2N_{\rm wt} - N_{\rm mt1}$, $M_{\rm mt2} = N_{\rm wt} - N_{\rm mt2}$, and $M_{\rm wt} = N_{\rm mt1} + N_{\rm mt2} - 3 N_{\rm wt}$, taking into account the multiplicity of amino acid changes and assuming a uniform point mutation rate, $\mu = 1.2 \times 10^{-5} \, {\rm day}^{-1}$~\cite{Neher2017} (SI Appendix). These dynamics determine the time-dependent viral load, $N(t) = \sum_i N_i (t)$, as well as the evolution of strain frequencies, $y_i (t) = N_i (t) / N(t)$. 

At a given antibody dosage, the strain dynamics of Eq.~[\ref{eq:sub_pop}] leads to a mutation-selection equilibrium that is dominated by the fittest strain (in the present system, we will show that this can be either wt, mt1, or mt2, depending on dosage). The equilibrium state has an infective viral load $\bar N_u$ and a total viral load $\bar N$ given by 
\EQ
\bar N_u (A) = \frac{\bar N (A)}{1+ A/K^*} = \frac{1}{C} \log \frac{f^*(A) + d}{d}, 
\label{eq:K}
\EE
up to mutational-load corrections of order $\mu$. Here $f^* (A) = \max_i f_i (A)$ is the dosage-dependent maximum fitness in the absence of saturation effects, and $K^*$ is the antibody resistance of the corresponding strain. The equilibrium viral load, which is the carrying capacity of HIV in its host- and dosage-specific environment, increases slower than linearly with the fitness $f^* (A)$. Given growth limitation by immune suppression, an increased fitness $f^*(A)$ is accompanied by a supra-linear increase in activation of the immune system that curbs viral replication (SI Appendix). In the following, we will show that the viral escape data can be described by the eco-evolutionary dynamics of Eqs.~[\ref{eq:birth_rate} -- \ref{eq:sub_pop}] with universal fitness parameters $b_i^0$, $K_i$, $d$, and 
host-specific niche constraints $C$. We will also compare our fitness model to an alternative model with additive niche constraint, $f_i (A, N) = f_i (A) - \tilde C N$, which has a linear carrying capacity, $\bar N (A) = f^* (A) / \tilde C$.

\paragraph{Viral load trajectories.} 
A convenient starting point for data analysis is the time-dependent viral load data under bnAb treatment in 11 hosts (Fig. 1A). Prior to the start of treatment, there is a host-specific load $N_0$, which we identify with the carrying capacity of HIV in a drug-free environment. In our model, the variation of initial loads translates into host-specific constraint parameters $C$, as given by Eq.~[\ref{eq:K}] (Table~S3). Following the infusion at time $t_0$, the antibody concentration decays exponentially,
\EQ
A(t) = A_0 \, \exp \left (-\frac{t - t_0}{\tau} \right )
\EE
with host-specific initial values $A_0$ in the range ($10^3 -10^5)$ $\mu$g/ml and a characteristic decay time $\tau = 10 \pm 3$ days (Fig.~S1, Table S2). In contrast, the relative load shows a common initial response to bnAbs, $N(t) / N_0 = \exp [-d_{\rm wt} (t - t_0)]$, from which we infer a universal clearance rate of the wild-type strain, $d_{\rm wt} = (0.45 \pm 0.03)$ ${\rm day}^{-1}$ (Fig.~1B). 
Consistently, antibody concentrations in the initial time interval are well above the average wild-type IC50 concentration $K_{\rm wt} = 0.08 \, \mu$g/ml~\cite{Caskey2017}, which implies that most virions are bound to antibodies and cannot replicate. The rebound of the viral load is driven by the bnAb-resistant strain mt1 in 9 of 11 hosts (see the data points in Fig.~3A). 
In these hosts, shifting the time axis to a common initial load, we obtain a data collapse describing the rebound to $\sim 90$\% of the initial load (Fig.~1C). This indicates a universal growth pattern that extends throughout the approach to the carrying capacity, as given by Eqs.~[\ref{eq:sub_pop}] and~[\ref{eq:K}]. The rebound is marked by an initially exponential increase of the load, $N(t) / N_0 \simeq x_{0, {\rm mt1}} \exp [f_{\rm mt1}^0 (t - t_0)]$, from which we estimate a universal basic growth rate $f^0_{\rm mt1} \equiv b^0_{\rm mt1} - d$ (we will show below that the growth of mt1 is independent of $A$ since $K_{\rm mt1} \gg A_0$). On the other hand, by extrapolating the measured (unshifted) rebound curves back  to $t = t_0$, we can infer the frequency $x_{0, {\rm mt1}}$ in the unperturbed viral population at the start of the treatment (Fig.~1D). We find host-specific mutant frequencies of order $10^{-3}$ to $10^{-2}$, in broad agreement with deep sequencing data of initial viral populations available in part of the individuals. This spread reflects strong environmental fluctuations governing the low-frequency dynamics.
To summarize, the observed load dynamics is consistent with universal fitness parameters and host-specific initial conditions of load and genetic variation.

\paragraph{Bayesian inference of the fitness model.}
To infer the full fitness model, we use a Bayesian procedure based jointly on the time series data of viral load and strain frequencies. Given a set of host-independent fitness parameters and host-dependent initial data, we compute trajectories of wild-type and mutant strains from Eq.~[\ref{eq:sub_pop}]. We evaluate the joint likelihood of the observed data across all individuals as a function of the model parameters, using a Gaussian error model (SI Appendix). To infer optimal parameters, we use a Markov Chain Monte Carlo algorithm that constructs a Markov chain, in which each state is a set of fitness parameters and initial mutant frequencies. The Markov chain converges to the posterior distribution of the fitness parameters; details are given in SI Appendix. 

The maximum-likelihood fitness model has strain-dependent basic replication rates $b_i^0$, resistance parameters $K_i$, and a strain-independent clearance rate $d$ (Table~S1). In Fig.~2A, the resulting basic fitness values $f^0_i = b^0_i - d$ in the absence of antibodies are plotted against $K_i$ for all strains; error bars indicate 95\% confidence intervals obtained from the posterior distribution. The universal maximum-likelihood fitness model reproduces the viral load trajectories across all 11 hosts (blue lines in Fig.~1A); that is, the considerable variability of these trajectories between hosts can be attributed to the variation in initial load $N_0$ (i.e. in niche constraint $C$) and in mutant frequency $x_{0, {\rm mt1}}$. In particular, the maximum-likelihood fitness parameters $f^0_{\rm mt1}$, $d$ and initial frequencies $x_{0, {\rm mt1}}$ agree well with the corresponding values obtained by fitting decline and rebound of the viral load (Table S1, Fig.~S2). 

Apart from estimating the fitness parameters in Eqs.~[\ref{eq:birth_rate}, \ref{eq:fitness}], Bayesian inference can also serve to rank the fitness model components against alternative functional forms. First, the data support the Michaelis-Menten function $b \sim (1 + A/K)^{-1}$ linking growth to antibody density; an alternative model with a fitness cost linear in $A$ has a significantly lower likelihood (SI Appendix). Second, the data favor the saturation model, $b \sim \exp(- CN_u)$, against an ecological model with linear niche constraint. The saturation model is also supported by the observed nonlinearity of the carrying capacity: the observed load ratio $\bar N_{\rm mt1} / \bar N_0 = 0.85 \pm 0.15$ is larger than the fitness ratio $f^0_{\rm mt1} / f^0_{\rm wt} = 0.5 \pm 0.12$ that would set the load ratio given a linear niche constraint (SI Appendix). 

\begin{figure*}
\centering
  \begin{adjustbox}{center}
   \includegraphics[height=45mm,width= 0.9\textwidth ]{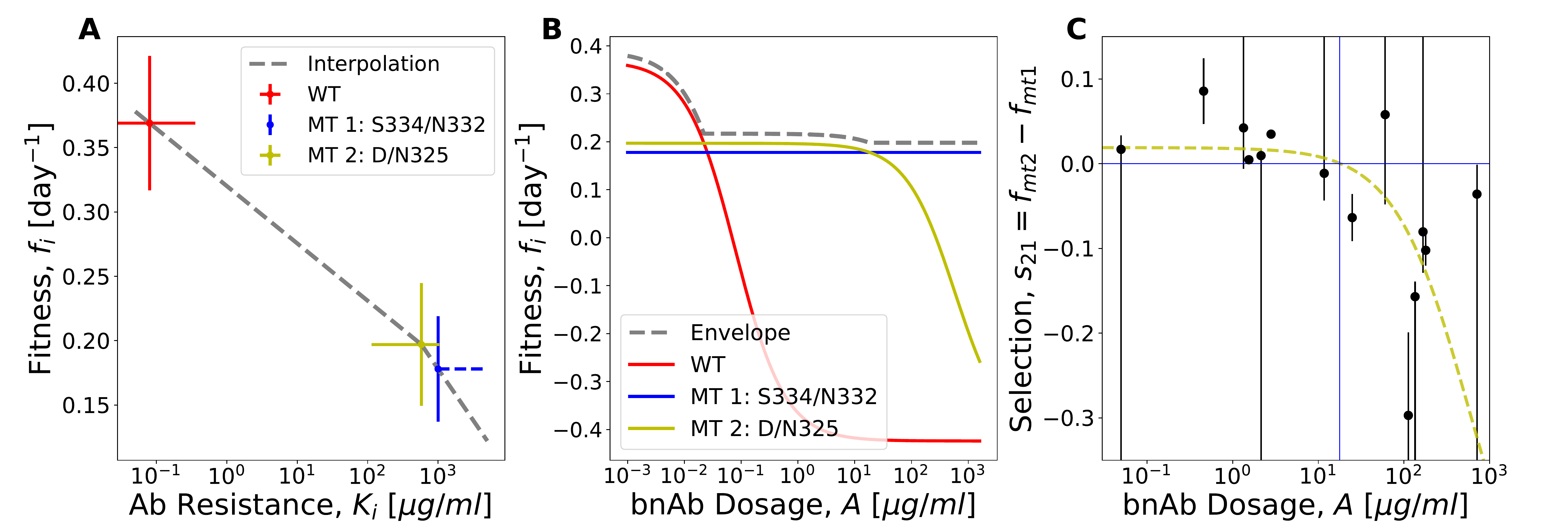}
 \end{adjustbox}
\caption{\small  
{\bf Resistance-cost tradeoff and dosage-dependent fitness ranking of viral strains.} 
(A)~The inferred drug-free growth rate, $f_i^0$, is plotted against the antibody resistance $K_i$ of the viral strains $i = $ wt, mt1, mt2 (maximum-likelihood values, error bars indicate 95\% confidence intervals). 
(B)~Michaelis-Menten growth profiles $f_i (A)$ of the strains $i = $ wt, mt1, mt2 interpolate between the basic growth rate $f^0_i$ and the clearance rate $d$ with an IC50 concentration $K_i$ (solid lines). The maximum growth rate $f^* (A)$ (dashed line) is the envelope of the growth profiles of individual strains. 
(C)~Dosage dependent selection between resistance mutants. Estimates of the selection coefficient $s_{21} = f_{\rm mt2} - f_{\rm mt1}$ obtained from relative frequency changes (dots, bars indicate sampling errors, large bars indicate inequalities involving frequencies below the sampling threshold) are compared to the predicted Michaelis-Menten form $\hat s_{21} (A)$ (yellow line). 
\label{fig3}
}
\end{figure*}

\paragraph{Resistance-cost tradeoff and dosage-dependent selection.}
As shown in Fig.~2A, the drug-free growth rate $f_i^0$ and the antibody resistance varies between strains in a correlated way: higher resistance, i.e., weaker binding to bnAbs, implies slower growth in the absence of antibodies. This resistance-cost tradeoff is intuitive: more drastic changes to a viral protein (here gp120) can more effectively reduce bnAb binding but also have larger impact on protein stability and/or reproductive functions, such as binding to host cells. Here we infer a tradeoff of considerable amplitude: the cost of escape, defined as $ (f^0_{\rm wt} - f^0_{\rm mt}) / f^0_{\rm wt}$, is about 50\% for both escape mutants. 

Fig.~2B shows the dosage-dependent growth profile $f_i (A)$ of the wild-type and escape mutant strains, as obtained from our maximum-likelihood fitness model. For a given strain, the growth rate takes a sigmoid (Michaelis-Menten) form that interpolates between the asymptotic values $f_i (A) \simeq f_i^0 = b^0_i - d$ in the low-binding regime ($A \ll K_i$) and $f_i (A) \simeq - d$ in the strong-binding regime ($A \gg K_i$) with a cross-over at the half-binding point (IC50 concentration, $A= K_i$). Specifically, mt1 is in the weak-binding regime throughout (i.e., $K_i \gg A_0$ in all hosts), and no dosage-dependence of its fitness is inferred from this data set. 

The growth profiles of Fig.~2B exhibit an important consequence of the resistance-cost tradeoff: the fitness ranking of viral strains depends on the antibody dosage. The fittest strain is the wild-type at low dosage, mt2 at intermediate dosage, and mt1 at high dosage. The resulting dosage-dependent maximum growth rate, $f^* (A)$, is the envelope of the growth profiles of individual strains (grey line in Fig.~2B); this growth rate determines a dosage-dependent carrying capacity by Eq.~[\ref{eq:K}]. Another aspect of the fitness ranking is a dosage-dependent selection coefficient between the two escape mutants, $s_{21} (A) = f_{\rm mt2} (A) - f_{\rm mt1} (A)$, which changes sign at a specific dosage $A_{12} \sim 10 \mu$g/ml. In Fig.~2C, we compare the model prediction $\hat s_{21} (A)$ (yellow line) with data points obtained from changes in frequency ratio between subsequent sampling time points, plotted against the average bnAb concentration in between these points (SI Appendix). The observed frequency changes are seen to be in agreement with the predicted functional form (large error bars reflect instances where at least one of the mutants is observed at only one of the two time points, resulting in an inequality for the corresponding data point for $s_{21}$). 

\begin{figure*}
\centering
  \begin{adjustbox}{center}
   \includegraphics[height=75mm,width= 0.95 \textwidth ]{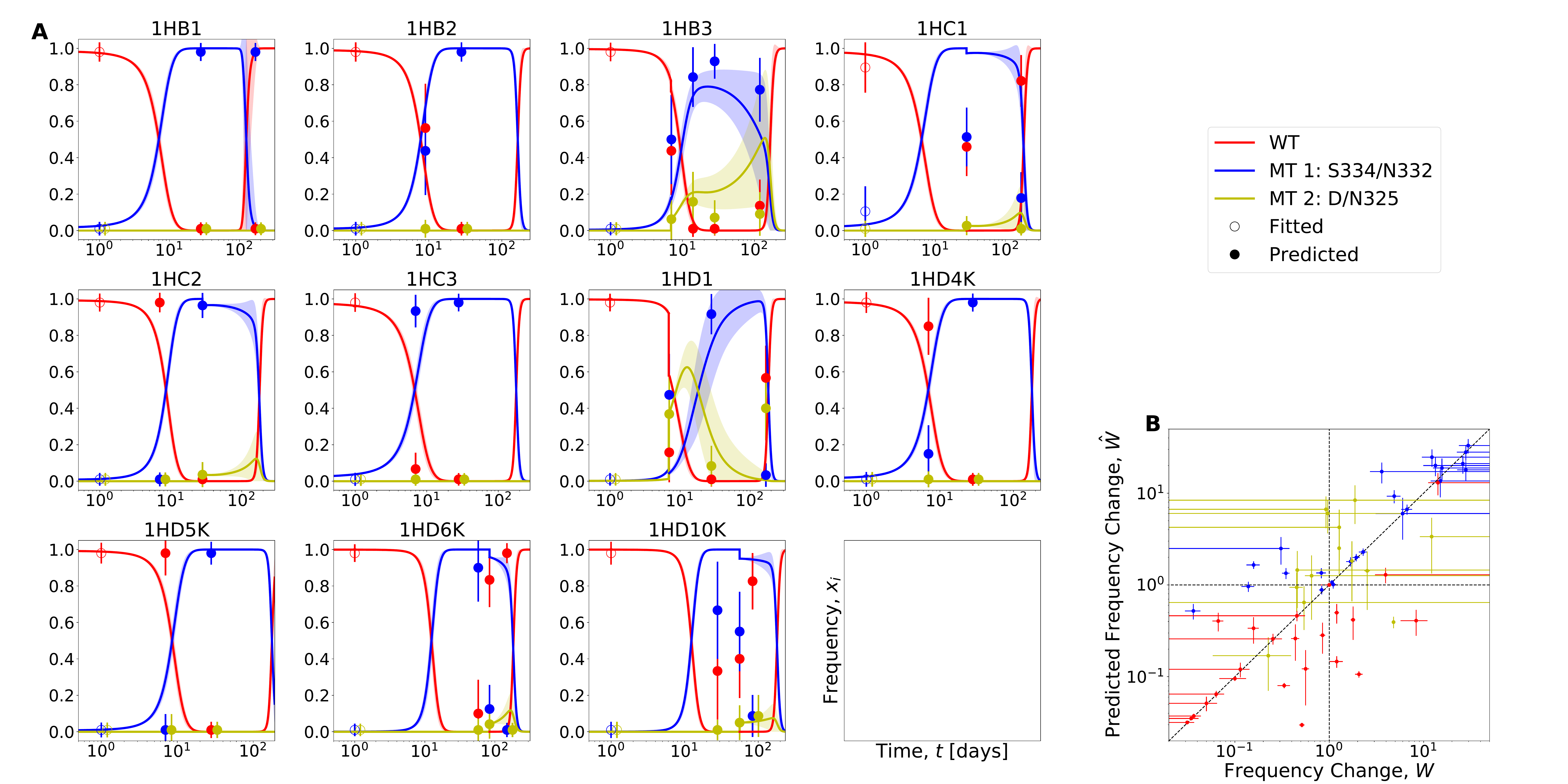}
 \end{adjustbox}
\caption{\small  
{\bf Prediction of escape evolution.} 
(A) Observed time series of strain frequencies (dots, bars indicate sampling errors) are shown together with predicted frequency trajectories for 11 validation protocols (lines). Fitness parameters used for predictions are obtained from complementary training sets. The first data point for each strain (open circles) is used as initial condition, the subsequent points (filled dots) are to be compared with predictions. When mt2 is first observed, the normalization of the predicted trajectories is updated. 
(B) Model predictions of strain frequency changes, $\hat w = \hat y_i (t_{k+1}) / y_i (t_{k})$ are plotted against the corresponding observed changes, $w = y_i (t_{k+1}) / y_i (t_{k})$. Frequency ratios with $y_{k+1}$ ($y_k$) below the sampling threshold are evaluated with pseudocounts. Frequency increase is correctly predicted in 22 of 28 instances (first quadrant), frequency decline in 21 of 28 instances (third quadrant). 
}
\label{fig2}
\end{figure*}

\paragraph{Predicting evolutionary escape trajectories.}
The resistance-cost tradeoff and the dosage-dependent fitness ranking of strains lead to an evolutionary prediction of our fitness model: in a bnAb treatment with time-dependent dosage, there is a reproducible turnover between prevalent strains. Here we test this prediction by training the model on subsets of 10 hosts and using frequency trajectories of the 11th host for prediction and validation. The host-specific initial data for the validation protocol are obtained as follows. The constraint coefficient $C$ is computed from the observed initial load by Eq.~[\ref{eq:K}]. The initial frequency $x_{0, {\rm mt1}}$ is obtained from sequencing data where available (in patients 1HB3, 1HD1), otherwise from backward extrapolation of the load rebound with a fitness parameter $f^0_{\rm mt1}$ inferred from the training set (Fig.~1C); we set $x_{0, {\rm mt2}} = 0$.


In Fig.~3A, we show the predicted frequency trajectories given by Eq.~[\ref{eq:sub_pop}] for 11 validation protocols together with the observed frequency data, which have been excluded from the model training step. The model successfully predicts salient qualitative features of the evolutionary strain turnover. First, the rebound of the viral load, which takes place while bnAb dosage is still high, is predominantly carried by mt1. Exceptions to this pattern are observed in hosts 1HC1 and 1HD10K, where the wild-type is not fully suppressed in the initial treatment phase, suggesting increased values of $K_{\rm wt}$ and a limitation of fitness universality. Second, at intermediate times and bnAb dosages, mt2 appears and qualitatively follows the predicted trajectories. However, we do not attempt to predict the appearance of mt2 in the observable frequency range; this would require a direct observation of its initial frequency $x_{0, {\rm mt2}}$. When mt2 appears, this observation is integrated into the initial data for predictions of the subsequent time points (the corresponding renormalization of frequencies is seen as jumps of the model trajectories). Third, a rebound of the wild-type occurs at late times and low bnAb dosages. This rebound is observed in almost all protocols where strain frequencies have been tracked over a sufficiently long time interval ($\gtrsim 50$ days). 

To test the quantitative predictability of frequency trajectories, we predict validation trajectories from one sampling time point to the next and compare model predictions of strain frequency changes, $\hat w = \hat y_i (t_{k+1}) / y_i (t_{k})$, with observed changes in the same time interval, $w = y_i (t_{k+1}) / y_i (t_{k})$ (Fig.~3C; hats indicate predicted quantities). The model correctly predicts 22 of 28 instances of frequency increase and 21 of 28 instances of decline, which amounts to an overall prediction accuracy of 77\%. The main limiting factor of predictability appears to be the incomplete knowledge of low-frequency mutations, which can be mitigated by deeper sampling of the initial populations.

\section*{Discussion} 
In this paper, we have established a fitness model for HIV-1 that predicts the {\em in vivo} eco-evolutionary response of the viral system to broadly neutralizing antibodies, given simple treatment protocols extending over limited periods. The key to predictability is the universality and simplicity of the fitness model, which operates on few, host-independent parameters. These are the intra-host replication rate, which depends on antibody density and viral population density, and the clearance rate, which is common to all strains. Recent mechanistic and genomic fitness models for HIV~\cite{Bonhoeffer2012, Ribeiro2013, Louie2018} provide a more detailed picture of the intra-host reproductive dynamics and the viral genome sites relevant for resistance evolution. In contrast, our model provides a coarse-grained description of the viral dynamics geared to inferability from limited {\em in vivo} time series data. 

Despite its simplicity, our fitness model has to incorporate the full immunological response of human hosts and the genetic variation of intra-host viral populations, which are key differences to {\em in vitro} replication assays. Hence, the (approximate) universality of fitness effects suggests biological features of antigen-antibody interactions. First, the resistance mutations analyzed in this study operate largely independently of their genetic background, which differs between viral populations in different hosts. It remains to be shown whether other bnAbs targeting more complex viral epitopes, such as the CD4-binding site, generate sizeable epistatic effects for resistance evolution that limit universality. Second, the host-to-host variation in immune response can, over the limited time intervals of our treatment protocols, be absorbed into a host-dependent carrying capacity, which acts as initial condition for the escape dynamics. Despite these simplifications, treatment protocols show a complex strain turnover, which is generated by time-dependent bnAb dosages together with a dosage-dependent fitness ranking between strains. This interplay is likely to be generic: bnAb treatment protocols generate antigenic fitness seascapes driving time-dependent strain prevalence. Given bnAb exposure over longer time scales, the strain dynamics is expected to become even more complex. In particular, compensatory mutations can stabilize mutant strains by reducing the fitness cost of antibody escape \cite{Gong2013, Lynch2015}. 

The key fitness characteristic driving viral escape evolution is a tradeoff between resistance against bnAb neutralization and growth in the absence of antibody challenge (Fig. 2A). This finding has implications for the optimization of bnAbs, which is a topic of high current interest \cite{Shaffer2016, Sprenger2020, Sachdeva2020}. The standard procedure to measure the power of bnAbs is based on neutralization assays against a panel of reference strains, which determine the potency (i.e., the mean IC50 against susceptible strains) and the breadth (i.e., the number of reference strains against which the IC50 is lower than a threshold value) of neutralization. However, these measures do not take into account the specific genetic changes that carry resistance evolution against that antibody. In contrast, the {\em in vivo} resistance-cost tradeoff provides an evolutionary measure of bnAb performance that integrates the effects of common resistance mutations. Given a Michaelis-Menten dependence of viral growth on antibody concentration and antibody resistance, the tradeoff function determines the maximum growth of resistant strains at a given bnAb dosage, $f^* (A)$ (Fig.~2B). This sets a necessary criterion for suitable bnAbs: at therapeutically sustainable dosages $A$, the growth of common resistance mutants must be suppressed, $f^* (A) < 0$. Importantly, resistance-cost tradeoffs and maximum-growth curves can be used to compare the performance of bnAbs targeting different viral epitopes, including antibody combinations. 

The predictability of the viral escape dynamics is also an important tool for designing optimal bnAb dosage protocols. Any such protocol has to balance medical limitations and physiological collateral costs of treatment with their effect of curbing the viral load. In particular, the effectivity of multi-step protocols crucially depends on timing and strength of individual applications~\cite{dasThakur2013, Mustonen2020}. The optimization of protocols has to take into account the full eco-evolutionary response of the viral population, including the turnover in strain prevalence. Here we have established a proof of principle that this response is computable for realistic treatment settings. Extending the method to other broadly neutralising antibodies and to more complex, resistance-limiting therapies with multiple bnAbs is an important avenue for future work.

\paragraph{\small Acknowledgements.} 
We acknowledge discussions with Armita Nourmohammed and Colin Lamont. 
This work has been supported by Deutsche Forschungsgemeinschaft grant SFB 1310.


\clearpage

\setcounter{equation}{0}
\setcounter{figure}{0}
\setcounter{table}{0}
\setcounter{page}{1}
\setcounter{section}{0}
\makeatletter

\renewcommand{\thefigure}{S\arabic{figure}}
\renewcommand{\theequation}{S\arabic{equation}}
\renewcommand{\thetable}{S\arabic{table}}

\section*{Methods}

\subsection*{Bayesian inference of the eco-evolutionary dynamics}

Here we describe the general method to infer the dynamics of a multi-strain pathogen population from time-series data of the total population size and of strain frequencies in the total population. As appropriate for the present data set, the method can be applied in a time-dependent and heterogeneous environment (here characterized by time-dependent antibody dosages and differences between human hosts), and it works with scarce frequency data (here, low initial frequencies of mutant strains are not observed and have to be inferred). Specifically, the method uses the following input data:
\begin{enumerate}[(a)]
\item time-series data of the total population size (viral load), $\NN$, consisting of measurements $N^\alpha (t_k) = \sum_i N_i^\alpha (t_k)$ in a series of host individuals $\alpha$ at different times $t_k$ ($k = 0, 1, \dots, K$). 
\item time-series data of strain frequencies, $\YY$, consisting of measurements $y_i^\alpha (t_k) = N_i^\alpha (t_k) / N^\alpha (t_k)$ for all strains~$i$ in the population. Frequency estimates are obtained from counts in a population sample, $y_i^{\alpha}(t_k) = n_i^{\alpha}(t_k) / N^{\alpha}_{\rm sample}(t_k)$, where $N_{\rm sample}^{\alpha}(t_k)$ is the sample size at time point $t_k$ in individual $\alpha$. Frequency data are available for a subset of the time points $t_k$ ($k = 0, \dots, K$). 
\item time-series data $\AA$ of an environmental parameter, here host-specific antibody dosages $A^\alpha (t_k)$. 
\end{enumerate}
We use these data to infer 
\begin{enumerate}[(a)]
\item the pathogen fitness model governing the eco-evolutionary dynamics of the multi-strain population. Here we use a primary fitness model of the form of Eqs. 1-3 in the main text. This model is specified by a set of parameters $F$ that includes strain-dependent basic replication rates $b_i^0$ and antibody resistances $K_i$, as well as a strain-independent clearance rate $d$. The fitness model is taken to be universal; that is, the fitness parameters are constrained to be uniform across all host environments $\alpha$. Alternative fitness models are discussed below. 
\item missing strain frequency data. Specifically, we infer a set of host-specific initial frequencies $X$ with entries $x^\alpha \equiv y_{\rm mt1}^\alpha (t_0), y_{\rm mt2}^\alpha (t_0) $, which are relevant as initial data for the strain dynamics. 
\item a set $C$ of host-dependent niche constraint parameters $C^\alpha$.
\end{enumerate}
In a Bayesian framework, we compute the posterior distribution of model parameters for a given data set. 
This inference method uses time series data $\NN, \YY, \AA$ from a full data set; a variant of the method for predictions across host environments will be discussed below. We now describe the evaluation of the Bayesian posterior. 

\paragraph{Pathogen population trajectories.}
Given the set of fitness parameters $F$, as well as the host-specific initial frequencies $X^\alpha$, constraint parameters $C^{\alpha}$, and antibody dosage measurements $\AA^\alpha$, the pathogen population dynamics given by Eqs. [1-3] can be solved using a simple Euler-step algorithm. This provides model-based 
\begin{enumerate}[(a)]
\item strain-specific population trajectories $\hat \NN_i^\alpha$ with values $\hat N_i^\alpha (t_k)$ at time points $t_k$ ($k = 0, \dots, K$), 
\item load trajectories $\hat \NN^\alpha$ with values $\hat N^\alpha (t_k) = \sum_i \hat N_i^\alpha (t_k)$,  
\item strain frequency trajectories $\hat \YY_i^\alpha$ with values $y_i^\alpha (t_k)= \hat N_i^\alpha (t_k) / \hat N^\alpha (t_k)$. 
\end{enumerate}

\paragraph{Likelihood of an eco-evolutionary model.} 
We evaluate the observed pathogen population trajectories under a given model by a log likelihood score additive in host environments and time points,
\EQA
\Sigma ({\rm data} \, | \, {\rm model} ) & = & \log P(\NN, \YY \| F, X, C, \AA)
\nonumber \\
& =& \sum_\alpha \sum_{k = 0}^K \left [ S_N (N^\alpha (t_k) \| \hat N^\alpha (t_k)) + S_Y (Y^\alpha (t_k) \| \hat Y^\alpha (t_k) ) \right ]. 
\label{Sigma} 
\EEA
The underlying error model is given by the score functions 
\EQA
S_N (N \| \hat N ) & = & - \frac{(\log N - \log \hat N)^2}{\sigma^2}
\\
S_Y (Y \| \hat Y ) & = & \log \left ( \frac{N_{\rm sample} !}{\prod_i n_i!}\prod_i \hat y_i^{n_i} \right ),
\EEA
which capture measurement errors of the load and sampling variation of frequencies, respectively. Using a flat prior distribution $P_0 (F, X, C)$, the likelihood score determines the Bayesian posterior distribution of model parameters, 
\EQ
\Sigma = \log P(F, X, C \| \NN, \YY, \AA).
\EE

\paragraph{Numerical evaluation of the score.}
We compute the likelihood score by a Monte Carlo Markov Chain (MCMC) algorithm as follows. At each step of the algorithm, a new set of fitness parameters and mutant frequencies is proposed: $F, X \rightarrow F', X'$. We solve Eqs. 1-3 for each individual $\alpha$ to obtain the trajectories $\hat{N}_i^{\alpha}$ and compute the new total likelihood $P(\NN, \YY | F', X', C, \AA)$. The new proposed parameters are then accepted using a Metropolis acceptance rate:
\EQ
\gamma =  \min \left [ 1, \frac{P(\NN, \YY | F', X', C, \AA) q(F, X | F', X')}{P(\NN, \YY | F', X', C, \AA) q(F', X' | F, X)} \right ],
\EE
where the new parameter set $F', X'$ is updated in a single position $j$ compared to the old set. The function $q(F',X'|F,X)$ gives the probability of the updated parameter set and is normally distributed, 
\EQ
q(F',X'|F,X) \sim \exp \left [- \frac{(F_j' - F_j)^2}{\beta^2 F_j^2} \right], 
\EE
when the parameter set is updated in position $j$. The step size of each update, $\beta$, is chosen such that the acceptance rate in the Markov chain is between the 20 and 40 \%.

We use the approach by Gelman and Rubin \cite{Rubin1992si} to check for convergence of the Markov chain: six independent chains are started with the initial fitness parameters drawn from uninformed, uniform distributions, and the initial mutant frequencies drawn from lognormal distributions. The variance of the posterior distribution is then produced as a mix of the between-chain and within-chain variances. The potential scale reduction factor $R_c$ estimates the potential decrease in the between-chains variability with respect to the within-chain variability. The factor is a measure of the convergence of the Markov chains. The reported values are obtained from the last $8 \cdot 10^5$ states of the Markov chain, after we let the chain converge for the same number of steps.

\subsection*{Application of the inference procedure}

\paragraph{Time-resolved {\em in vivo} data of viral load and strain frequencies.}
Our analysis of bnAb escape is performed on data from the study of Marina Caskey \emph{et al.}~\cite{Caskey2017si}, which includes both time-series data of the total population size $\NN$ and time-series data of strain frequencies $\YY$. This study investigated the effectiveness of the broadly neutralizing antibody (bnAb) 10-1074 in a group of 19 individuals with HIV-1 infection. Here we exclude individuals that were on antiretroviral therapy (3 individuals), were already infected with 10-1074 resistant mutants (2 individuals), or had single genome sequencing at less than 3 time points (3 individuals), leaving a set of 11 individuals for further analysis. We use the following data reported in ref.~\cite{Caskey2017si}:
\begin{enumerate}[(a)]
\item The time-dependent bnAb concentration $A (t)$  in blood plasma shows an exponential pattern with characteristic decay time $\tau = (10 \pm 3)$ days (mean, 95\% confidence interval) approximately uniform across patients; see Fig. \ref{fig:raw_data}. The initial bnAb dosage $A_0$ varies in the range $10^2$ to $10^3$ $\mu$g/ml, following a single infusion of bnAb 10-1074 at dosages of 10 or 30 mg/kg (Table~S2). 

\item The viral load $N^\alpha (t)$ is recorded at an average of 12 time points after the bnAb infusion; measurements have a log error of 0.2 or 0.3. The initial viral load $N_0$ varies in the range $10^3$ to $10^5$ virions/ml (Table~S2). 

\item Single-virion genome sequencing data produce frequency estimates for the sensitive strains (containing the wt allele at positions gp-120 334, 332, and 335) and the resistance strains mt1 (containing an amino acid change at position 334 or 332) and mt2 (containing an amino acid change at position 335). Frequency data are available at 3 to 5 time points with an average sample size $N_{\rm sample} = 24$. 

\item The neutralization sensitivity sensitive strains is measured in terms of their IC50 concentration, with an average value $K_{\rm wt} = 0.08$ across all individuals.  This value is used as an input to our inference procedure. 
\end{enumerate}

\paragraph{Inference of the fitness model.}
Given this input data, our Bayesian inference procedure yields maximum-likelihood fitness parameters entering (Table S1); these parameters enter the growth rates of Eqs.~[1--2]. The confidence interval for each parameter is obtained from the posterior distribution of the Markov chain. Our inference uses a score function of the form (\ref{Sigma}) with additional prefactors $c_N (t_k)$ and $c_Y (t_k)$ for the score components $S_N$ and $S_Y$, respectively. These factors upweigh earlier time points (with larger load and frequency gradients) over later time points and frequency data (which are available at fewer time points) over load data. 

The reported fitness parameters have converged to their posterior distributions, as indicated by the scale reduction factor $R_C$, which is below $1.05$ for the replication and clearance rates but above $1.3$ for the antibody resistance measures $K_{\rm mt1}$, $K_{\rm mt2}$. The antibody resistance $K_{\rm mt1}$ did not converge to any finite value; that is we do not find evidence that mt1 strains are affected by the antibodies in the range of bnAb dosages used in ref.~\cite{Caskey2017si}. This is corroborated by the experimental observation of a loss of the bnAb binding site in this mutant. In contrast, we find an intermediate resistance $K_{\rm mt2} \sim (6 \pm 4) \times 10^2 \mu$g/ml (maximum likelihood value, 95\% confidence interval). Thus, mt2 confers an intermediate level of resistance, leading to reduced reproduction rates at large bnAb concentrations. Because of the few data points and uncertainties of the frequency of mt2 at late times, the inferred value $K_{\rm mt2}$ has a sizeable error margin. The host-specific  constraint parameters $C^{\alpha}$ are not inferred independently, but are determined self-consistently from the initial load data and the fitness parameters, $
C^{\alpha} = \log [( f^0_{\rm wt} + d) / d] /N_0^\alpha$, as given by Eq.~[4].

\paragraph{Consistency of the inference procedure.}
As a consistency check for the full Bayesian inference, we can compare two of the maximum-likelihood fitness parameters, $f_{mt1}$ and $d$, with the corresponding values obtained from direct fits of the viral load data (Fig. 1BC). This comparison is reported in Table S1, showing good agreement. 

Additionally, we can compare the values for $x^{\alpha}_{\rm mt1}$ that we infer using the full data set with the values $x_{0, {\rm mt1}}$ that we obtain from Fig. 1D, using viral load data only. Figure \ref{fig:S2} compares both values (errors bars indicate 95\% confidence intervals). We find that both inference methods yield consistent results for the initial frequency of mutant 1.

\paragraph{Fitness model ranking.}
To test the specificity of our fitness model, we compare it to two alternative models with different fitness components. The first model stipulates a fitness cost linear in $A$, the second model load limitation by a niche constraint linear in the total viral load $N$. We compare fitness models by the Bayesian Information Criterion, ${\rm BIC} = k \log n - 2 \Sigma_{\rm max}$, where $n$ is the total number of data points, $k$ is the number of model parameters, and $\Sigma_{\rm max}$ is the maximum log likelihood score obtained from Eq.~(\ref{Sigma}). We consider the following models:

\begin{enumerate}[(a)]
\item {\bf Biophysical model with saturation of replication rate.} 
This model, which is used in the analysis of the main text, is characterised by two key assumptions: (1) The antibody dosage $A$ affects the viral replication rate by Michaelis-Menten kinetics, (2) the replication rate is limited by saturation effects that depend on the density of infective (bnAb-unbound) virions. The viral fitness takes the form 
\EQA
f_i (A, N_u) & = & b_i^0 \, \frac{\exp(-CN_u)}{1 + A/K_i}  - d, 
\nonumber \\
& \equiv & f_i (A) - \Delta f_i (A,N_u), 
\label{model1}
\EEA
as given by Eqs.~[1--2], where $f_i(A)$ is the asymptotic growth rate in the regime of low $N_u$. The term in the denominator describes the Michaelis-Menten dependence of the fitness on the antibody concentration, scaled in terms of the dissociation constant of functional (growth-neutralizing) binding. We use a Hill factor 1, assuming independent binding events. The exponential constraint factor $\exp(-CN_u)$ is an approximate form to describe the density dependent reduction of viral replication in infected cells. This reduction can be caused by the activation of cytotoxic (CD8+) T cells \cite{Boer1998si}. For a given host immune system, a larger effective viral load $N_u$ corresponds to an increased number of infected CD4+ T cells that give a stronger signal for the activation and recruitment of the immune response with cytotoxic T cells. In this case, the exponential factor $\exp(-CN_u)$ describes a Michaelis-Menten dependence of activation on recognition of viral peptides in the non-saturated regime. A similar reduction of viral replication can arise from a local depletion of \emph{uninfected} CD4+ T cells that can still be functionally infected \cite{Boer1998si}. In this case, the exponential factor results from a Poisson distribution of multiple infection events.

In the biophysical model, the impact of a maximum fitness $f^*(A)$ on the host can be characterized by two quantities. (i) The viral population saturates at the stationary viral load, which takes the form 
\EQ
\bar N (A) = \frac{1+ A/K^*}{C} \log \frac{f^*(A) + d}{d}, 
\EE
as given by Eq.~[4], where we assume that the population contains a dominant fittest strain with parameters $f^* (A), d, K^*$. (ii) Assuming that the reduction of viral growth $\Delta f_i (A, N_u)$ occurs by immune interactions involving cytotoxic T cells, the resulting activation of the immune system, $L$, can be quantified as, 
\EQ
L \sim \bar N_u (A) \, f^*(A). 
\EE
The stationary viral load depends in a sub-linear way, the immune system activation in a super-linear way on the fitness $f^*(A)$ of the viral population. 

\item {\bf Linear antibody fitness cost.} In this model, the viral fitness has a linear dependence on the antibody concentration, 
\EQA
f_i (A, N_u) & = & b_i^0  \left ( 1 -  \frac{A}{\tilde K_i} \right )\exp(-C N_u)   - d
\label{model2}
\EEA
where the coefficient $\tilde{K}_i$ is to be inferred for each variant. Compared to the biophysical growth model of Eqs.~[1--2], this model has $\Delta \Sigma_{\max} = -272$ and $\Delta k = 1$ (the parameter $\tilde K_{\rm wt}$ is to be inferred), giving a large information drop, $\Delta({\rm BIC}) =  550$. We conclude that the linear model does not accurately describe the escape dynamics from the bnAb, which provides evidence for Michaelis-Menten viral growth dynamics from {\em in-vivo} data. 

\item {\bf Linear ecological model.} The viral fitness decreases proportionally to the viral population size, 
\EQA
f_i (A, N_u) & = & \frac{ b_i^0 }{1 + A/K_i} - d - \tilde C N 
\nonumber \\
& \equiv & f_i (A) - \tilde C N 
\label{model3}
\EEA
with an ecological constraint parameter $\tilde C$ that is independent of the viral lineage. In contrast to the saturation model, the linear ecological model describes standard logistic growth up to a carrying capacity,  
\EQ
\bar N (A) = \frac{f^* (A)}{\tilde C}, 
\EE 
that is proportional to the replication rate $f^*(A)$. This difference can be traced in the observed eco-evolutionary pattern. The predicted load ratio between mutant and wild type strains is 
\EQ
\frac{\bar{N}_{\rm mt}}{ N_0} =  
\left \{
\begin{array}{ll} 
\dfrac{\log \left[ (f_{\rm mt} +d)/d \right]}{\log \left[(f_{\rm wt} + d) / d \right]} 
& \mbox{(saturation model)} 
\\ \dfrac{f_{\rm mt}}{f_{\rm wt}}
& \mbox{(linear ecological model).} 
\end{array} \right. 
\EE
The observed load ratio is $\bar{N}_{\rm mt} / N_0 \approx 0.85$ at a fitness ratio $f_{\rm mt} / f_{\rm wt} \approx 0.5$, supporting the sublinear pattern predicted by the saturation model. Consistently, the linear model has an information drop $\Delta({\rm BIC}) = - 2 \Delta \Sigma_{\max} = 24$ in the Bayesian inference scheme. Together, the {\em in-vivo} data support the saturation model. 
\end{enumerate}

\paragraph{Inference of dosage-dependent selection.}
In Fig.~2C, we infer a dosage-dependent selection coefficient
\EQ
s_{21}(A) = f_{\rm mt2} (A) - f_{\rm mt1} (A) 
\EE
between resistant strains with different escape mutations. The empirical selection coefficient data is obtained from the observed frequency trajectories $\YY_{\rm mt1}, \YY_{\rm mt2}$ as follows. The time-dependence of the log frequency ratio, $\xi (t) \equiv \log [y_{\rm mt2} (t) / y_{\rm mt1} (t)]$, is given by the instantaneous selection coefficient, 
\EQ
\frac{d}{dt} \xi (t) = s_{21} (A(t)), 
\EE
where we assume $y_{\rm mt1} + y_{\rm mt2} = 1$ (i.e., we ignore the wt strain) and consider only frequency changes by selection (i.e., we ignore mutational turnover).  
Hence, the observed log frequency change in a time interval $(t_{k}, t_{k+1})$ measures the average selection coefficient, 
\EQ
\frac{\xi (t_{k+1}) - \xi(t_k)}{t_{k+1} - t_k} = \bar s_{21} (t_{k+1}, t_k), 
\EE
which is plotted in Fig.~2C against the average average bnAb dosage in the same time window between the two frequency measurements. The frequencies $y_{\rm mt1}$ and $y_{\rm mt2}$ are obtained from the sample counts, $y_i = N_i / N_{\rm sample}$ with the following pseudocounts. When $N_i =0$ for mt1 at $t_k$ and/or for mt2 at $t_{k+1}$, the frequency is substituted by $y_i = 1/(N_{\rm sample}+1)$ and the resulting selection coefficient is recorded as a maximum estimate. When $N_i =0$ for mt1 at $t_{k+1}$ and/or for mt2 at $t_k$, the frequency is again substituted by $y_i = 1/(N_{\rm sample}+1)$ and the selection coefficient is recorded as a minimum estimate.

\subsection*{Predicting viral escape evolution}

To test the predictive power of the fitness model across different host environments, we employ a different inference scheme that partitions the data into disjoint training and validation sets. Specifically, we compute predicted frequency trajectories $\hat \YY^\alpha$ in a given individual $\alpha$ based on fitness model parameters inferred from the hosts excluding $\alpha$, and we compare these predictions to the observed trajectories $\YY^\alpha$. We use two variants of the validation scheme: (1) We compute full frequency trajectories $\hat \YY^\alpha$ using the initial mutant frequency $x^\alpha_{0, {\rm mt1}}$ obtained from viral load data, (2) we evaluate frequency changes between subsequent sampling points, predicting $\hat Y^{\alpha}(t_{k+1})$ with $Y^{\alpha}(t_{k})$ as input data. The prediction and validation procedure consists of the following steps. 

\paragraph{Prediction and validation of frequency trajectories.}
In order to predict the escape evolution for a given individual $\alpha$, we use the following scheme:
\begin{enumerate}[(a)]
\item Data partitioning. We divide the data into a training set $(\mathbf{N}, \mathbf{Y}^{-\alpha})$ that contains the frequency data of all individuals $\beta \neq \alpha$ and a test set $\YY^\alpha$ that contains the frequency data of individual $\alpha$. 

\item Fitness parameters. Growth and resistance parameters $b_i$, $d$ and $K_i$ are inferred from $(\mathbf{N}^{-\alpha}, \mathbf{Y}^{-\alpha})$, as described above (Fig.~S3). 

\item Host-specific data. Initial load $N_0^\alpha$ and initial dosage $A_0$ are given by measurements, the initial frequency $x^\alpha_{\rm mt}$ is inferred from $\NN^\alpha$ (see Fig.~1C), the constraint parameter is given by $C^{\alpha} = \log [( f^0_{\rm wt} + d) / d] /N_0^\alpha$; see Eq.~[4]. 

\item Prediction of population trajectories. We compute trajectories $\hat \NN^\alpha = (\hat{\NN}^\alpha_{\rm wt}, \hat{\NN}^\alpha_{\rm mt1}, \hat{\NN}^\alpha_{\rm mt2})$ from Eqs.~[1--3], using the fitness parameters obtained from step 2 and the environmental parameters obtained from step 3. 

\item Prediction and validation of frequency trajectories. The predicted trajectories $\hat{\YY}^{\alpha}_i=\hat{N}_i / \hat{N}_{\rm tot}$ with values $y_i^\alpha (t_k)= \hat N_i^\alpha (t_k) / (\sum_i \hat N_i^\alpha (t_k))$ are compared to the observed frequency trajectories $\YY^\alpha$. Trajectories for mt2 are predicted only for times after its first appearance (at a time $t^\alpha_{\rm mt2}$); the normalization of all frequencies is adjusted for the predictions at subsequent time points. 

\end{enumerate}
Repeating this procedure for every individual in the data set, we obtain the 11 predictions shown in Fig. 3A. In this scheme, frequency predictions for individual $\alpha$ at time $t_{k+1}$ use frequency data of that individual only at $t_0$ (wt, mt1) and at $t^\alpha_{\rm mt2} \leq t_k$ (mt2) as input. 

\paragraph{Prediction and validation of frequency changes.}
To evaluate escape evolution over time intervals $(t_k, t_{k+1})$ between subsequent sampling points, we use a variant of the prediction/validation scheme. We use available data $(N^\alpha (t_k), Y^\alpha (t_k))$ as initial data and compute $\hat y_i^\alpha (t_{k+1})= \hat N_i^\alpha (t_{k+1}) / (\sum_i \hat N_i^\alpha (t_{k+1}))$ from Eqs.~[1--3]. Next, we simulate the effects of sampling on the predicted frequencies, using a set of 1000 samples of size $N_{\rm sample}(t_{k+1})$. 
For a variant with population frequency $\hat y_i^\alpha (t_{k+1})$, we obtain the predicted mean sample frequency $\hat y_{i, {\rm sample}}^\alpha (t_{k+1})$ by averaging over 1000 sample counts $\hat N_{i, {\rm sample}}^\alpha (t_{k+1})$ and substituting samples with zero counts by pseudocounts (see above). The same procedure is applied to the observed  frequency data $y_i^\alpha (t_{k+1})$ used for validation. In Fig.~3B, we plot the predicted mean sample frequency changes $\hat{W} = \hat{y}_i(t_{k+1}) / y_i(t_k)$ against the corresponding observed changes $W = y_i(t_{k+1}) / y_i(t_k)$. This scheme allows for the most stringent test of the prediction method:   frequency predictions for individual $\alpha$ at time $t_{k+1}$ use data of that individual only at $t_k$ as input.

\paragraph{Universality of fitness parameters.}
We can compare the values of the fitness parameters $b_{\rm wt}$, $b_{\rm mt1}$, $b_{\rm mt2}$ as inferred from the full data set (Table S1) with the values that we get from the reduced training sets $(\mathbf{N}, \mathbf{Y}^{-\alpha})$. Fig.~S3 shows, for each parameter, the estimates from 11 training sets together with the inferred value from the full data set. The consistency of these inferences supports the approximate universality (i.e., host independence) of the basic fitness parameters. 


\vspace*{4cm}

\begin{table}[h]
\begin{center}
\begin{tabular}{ll|ll}
\label{tab:Fitness}
 fitness parameter && Bayesian posterior & fit value \\
 \hline
$b_{\rm wt}$ & $[{\rm day^{-1}}]$ & $0.79 \pm 0.06$ & $-$ \\
$b_{\rm mt1}$ & $[{\rm day^{-1}}]$& $0.60 \pm 0.04$ & $0.65 \pm 0.05$ \\
$K_{\rm mt1}$ & $[\mu {\rm g/ml}]$ & $\rightarrow \infty$ & $-$  \\
$b_{\rm mt2}$ & $[{\rm day^{-1}}]$ & $0.61 \pm 0.04$ & $-$ \\
$K_{\rm mt2}$ & $[\mu {\rm g/ml}]$ & $6 \cdot 10^2 \pm 4 \cdot 10^2$ & $-$ \\
$d$ & $[{\rm day^{-1}}]$ & $0.42 \pm 0.03$ & $0.45 \pm 0.03$ 
\end{tabular}
\end{center}
\caption{{\bf Universal fitness parameters.} Strain-specific basic reproduction rate, $b_i$, and strain-specific dissociation constant, $K_i$, for the strains $i =$ wt, mt1, mt2; clearance rate for all strains, $d$. Bayesian posterior values (maximum likelihood values, 95\% confidence interval) from the full data set are compared to direct fits to the viral load data (Fig.~1B,C). The inference procedure is detailed in SI Text.   }
\end{table}

\vspace*{1cm}

\begin{table}[h]
\begin{center}
\begin{tabular}{l|lllll}
\label{tab:host_specific}
 host, $\alpha$ & $A^\alpha_0 [\mu {\rm g/ml}]$ & $N^\alpha_0 [{\rm copies / ml}]$ & $x^\alpha_{\rm mt1}$ (Bayesian) & $x^\alpha_{0, {\rm mt1}}$ (fit) & $C^\alpha$\\
 \hline
 1HB1  & $1.4 \cdot 10^2$ & $3.0 \cdot 10^4$ & $0.011 \pm 0.008$ & $0.009 \pm 0.006$ & $4.5 \cdot 10^{-5}$ \\
 1HB2 & $2.2 \cdot 10^2$ &  $1.3 \cdot 10^4$  & $0.0045 \pm 0.003$ & $0.005 \pm 0.004$ & $9.9 \cdot 10^{-5}$\\
 1HB3  & $1.5 \cdot 10^2$ & $7.7 \cdot 10^4$ & $0.031 \pm 0.023$ & $0.025 \pm 0.012$ & $1.8 \cdot 10^{-5}$\\
 1HC1    & $5.0 \cdot 10^2$ & $3.1 \cdot 10^3$ & $0.048 \pm 0.041$ & $0.014 \pm 0.007$ & $4.3 \cdot 10^{-4}$ \\
 1HC2 & $7.0 \cdot 10^2$ & $1.7 \cdot 10^5$ & $0.003 \pm 0.002$ & $0.003 \pm 0.002$ & $8.1 \cdot 10^{-6}$ \\
 1HC3& $8.3 \cdot 10^2$ & $4.3 \cdot 10^4$ & $0.024 \pm 0.014$ & $0.017 \pm 0.015$ & $3.1 \cdot 10^{-5}$ \\
 1HD1   & $1.0 \cdot 10^3$ & $9.0 \cdot 10^4$ & $0.032 \pm 0.019$ & $0.053 \pm 0.02$  & $1.5 \cdot 10^{-5}$\\
 1HD4K& $7.8 \cdot 10^2$ & $1.3 \cdot 10^5$ & $0.008 \pm 0.005$  & $0.011 \pm 0.004$ & $1.0 \cdot 10^{-5}$\\
 1HD5K & $6.4 \cdot 10^2$ & $1.6 \cdot 10^4$ & $0.004 \pm 0.003$ & $0.003 \pm 0.002$  & $8.3 \cdot 10^{-5}$ \\
 1HD6K & $8.8 \cdot 10^2$ & $2.2 \cdot 10^4$  & $0.0003 \pm 0.0004$ & $-$  & $6.3 \cdot 10^{-5}$ \\
 1HD10K & $1.0 \cdot 10^3$ & $2.3 \cdot 10^4$ & $0.0003 \pm 0.0004$  & $-$ & $6.0 \cdot 10^{-5}$ 
\end{tabular}
\end{center}
\caption{{\bf Host specific model parameters.} Initial antibody dosage, $A^\alpha_0$, initial viral load, $N^\alpha_0$ (measured values), initial mutant frequency, $x^\alpha_{\rm mt1}$ (Bayesian posterior from full data set), initial mutant frequency $x^\alpha_{0, {\rm mt1}}$ (fit value from viral load data $\NN$), niche constraint parameter, $C^\alpha$ (Bayesian posterior from full data set). The inference procedure is detailed in SI Text. }
\end{table}

\begin{figure}[b]
\begin{center}
\includegraphics[width= 0.6\textwidth]{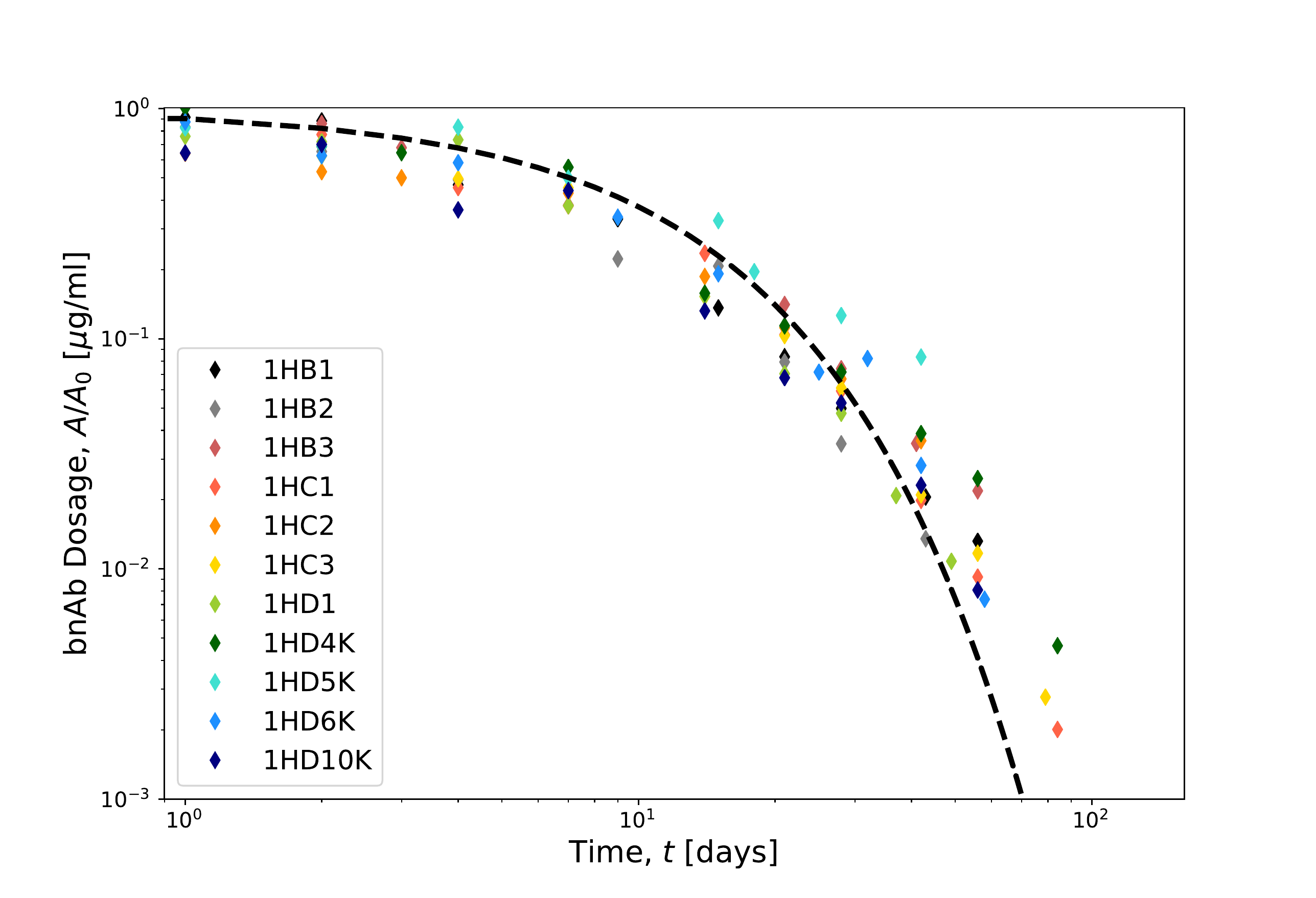}
\end{center}
\caption{\small  {\bf Time-dependent antibody dosage.} 
Measurements of the time-dependent dosage relative to the initial dosage, $A^\alpha(t)/A^\alpha_0$ (dots), for the 11 patients of this study are shown together with an exponential fit curve with characteristic decay time $\tau = 10 \pm 3$ days.
}
\label{fig:raw_data}
\end{figure}


\begin{figure}
\begin{center}
\includegraphics[width= 0.65\textwidth]{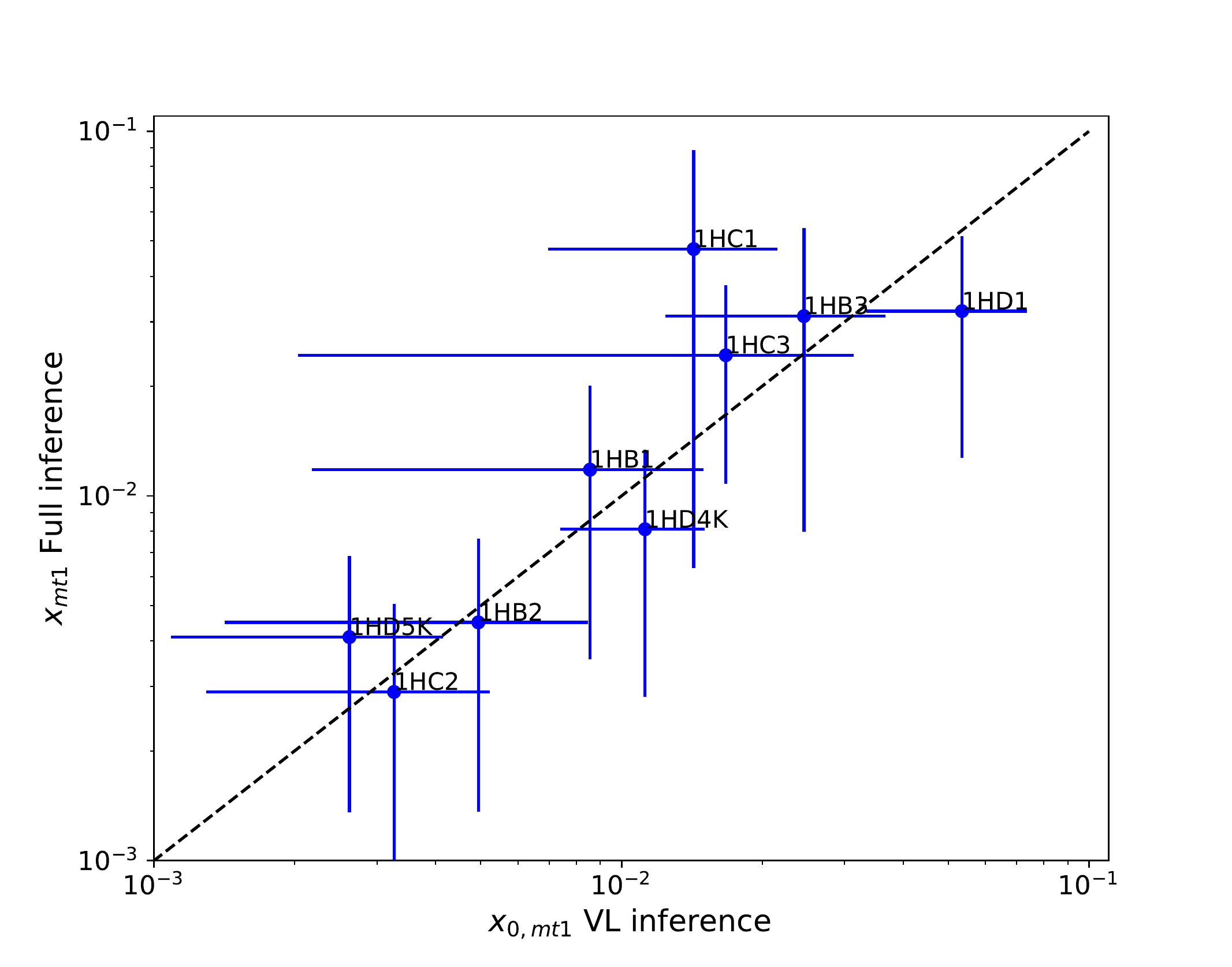}
\end{center}
\caption{\small  {\bf Consistency of inference procedures.} The host-specific initial mutant frequency $x^\alpha = y_{ {\rm mt1}} (t_0)$ (maximum-likelihood value, 95\% confidence interval) obtained by Bayesian inference from the full data set is plotted against the corresponding frequency obtained from direct fits of the viral load data (cf.~Fig.~1D; this value is used for predictions). }
\label{fig:S2}
\end{figure}

\begin{figure}[t]
\begin{center}
\includegraphics[width= 0.8\textwidth]{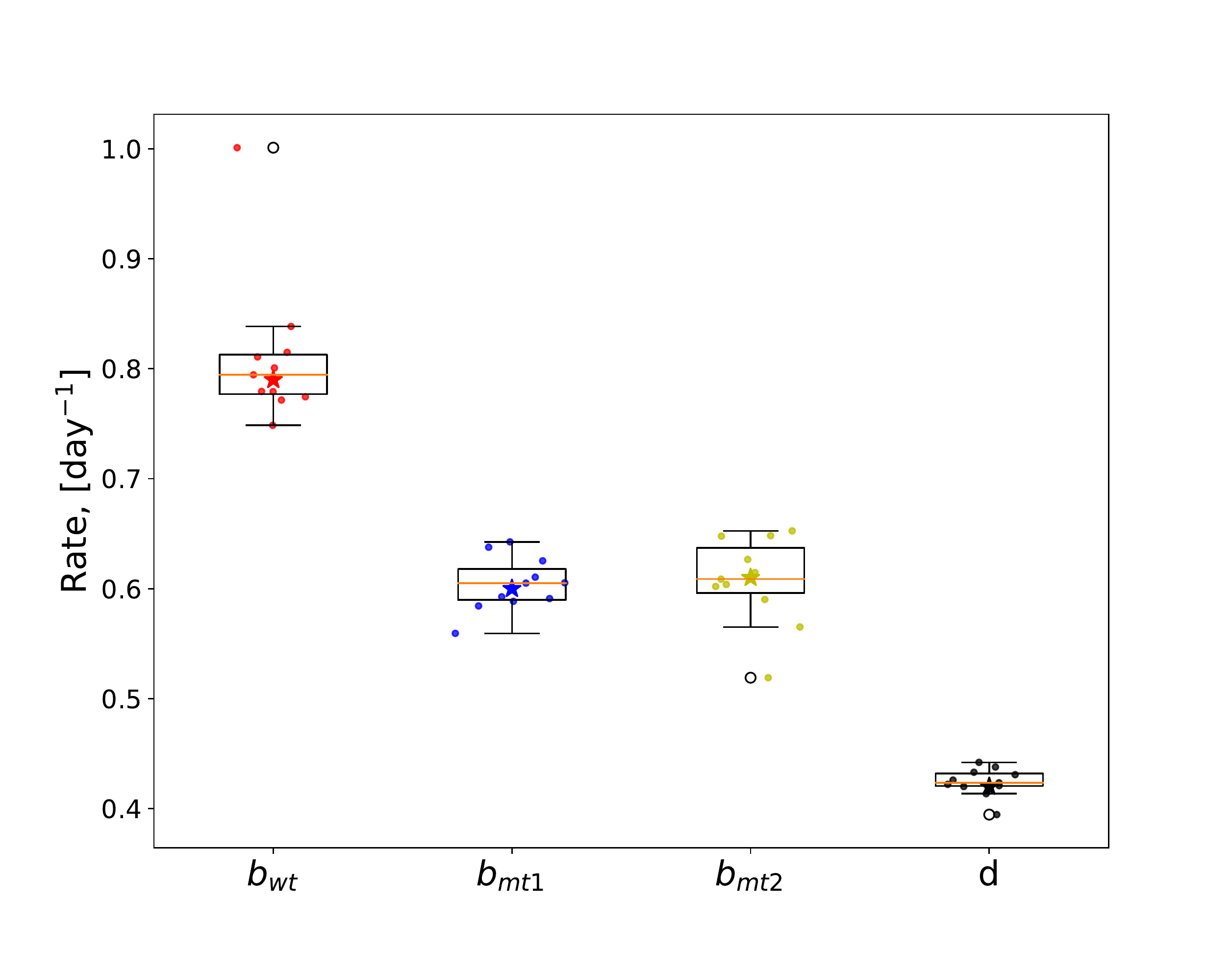}
\end{center}
\caption{\small  {\bf Universality of the fitness model.} Inferred maximum-likelihood fitness parameters $b_{\rm wt}$, $b_{\rm mt1}$, $b_{\rm mt2}$, and $d$ obtained from 11 training data sets of 10/11 hosts (dots; these values are used for predictions), box plots of these parameters, and corresponding parameter values inferred from the full data set (stars).
}
\label{fig:S3}
\end{figure}

\end{document}